\documentclass[a4paper]{jpconf}
\usepackage{graphicx}
\usepackage{amssymb,amsmath,bm,bbm}
\usepackage{cite}
\usepackage{color}
\usepackage{graphicx}
\usepackage{slashed} 

\begin{document}
\title{Dark matter conversion as a source of boost factor
for explaining the cosmic ray positron and electron excesses 
}

\author{Ze-Peng Liu, \ Yue-Liang Wu, \ and  Yu-Feng Zhou}

\address{
State Key Laboratory of Theoretical Physics,\\
Kavli Institute for Theoretical Physics China,\\
Institute of Theoretical Physics, \\
Chinese Academy of Sciences, Beijing, 100190, P.R. China
}

\ead{yfzhou@itp.ac.cn}






\begin{abstract}
  In interacting multi-component dark matter (DM) models, if the DM components
  are nearly degenerate in mass and the interactions between them are strong
  enough, the relatively heavy DM components can be converted into lighter
  ones at late time after the thermal decoupling. Consequently, the relic
  density of the lightest DM component can be considerably enhanced at late
  time. This may contribute to an alternative source of boost factor required
  to explain the positron and electron excesses reported by the recent DM
  indirect search experiments such as PAMELA, Fermi-LAT and HESS etc..
\end{abstract}


In the recent years, a number of experiments such as PAMELA
\cite{Adriani:2008zr}, ATIC \cite{:2008zzr}, Fermi-LAT \cite{Abdo:2009zk} and
HESS \cite{Aharonian:2009ah} etc.  have reported excesses in the high energy
spectrum of cosmic-ray positrons and electrons over the backgrounds estimated
from the standard astrophysics, which may be interpreted as indirect signals of
the annihilation or decay of dark matter (DM) in the Galactic halo. 
If the DM particles are  thermal relics such as the weakly interacting massive
particles (WIMPs), the thermally averaged product of their annihilation cross
section with the relative velocity at the time of thermal freeze out is typically
$\langle\sigma v\rangle_F\simeq 3\times
10^{-26}\mbox{cm}^{3}\mbox{s}^{-1}$. The positron or electron flux produced by
the DM annihilation can be parametrized by
\begin{eqnarray}
\Phi_{e} & = & B\overline{N}_{e}\frac{\rho_{0}^{2}\langle\sigma v\rangle_F}{m_D^{2}} ,
\end{eqnarray}
where $\rho_{0}$ is the smooth local halo DM
energy density estimated from astrophysics, $\overline{N}_{e}$ is the averaged
electron number produced per DM annihilation which depends on DM models and
parameters in the models for the propagation of cosmic ray particles, and $m_D$ is the mass of the DM particle. The boost
factor $B$ is defined as $B\equiv(\rho/\rho_{0})^{2}\langle\sigma
v\rangle/\langle\sigma v\rangle_F$ with $\rho$ the true local DM density and
$\langle \sigma v \rangle$ the  DM annihilation cross section multiplied 
by the relative velocity and averaged over the DM velocity distribution  today.
Both the PAMELA and Fermi-LAT results indicate that a large
boost factor is needed~\cite{Cholis:2008hb,Bergstrom:2009fa}. For a typical DM
mass of $\sim$1(1.6) TeV the required boost factor is $B\sim 500(1000)$ for
DM annihilating directly into $\mu^+\mu^-$ and $\rho$ fixed at
$\rho_0=0.3\mbox{ GeV}\cdot\mbox{cm}^{-3}$ ~\cite{Bergstrom:2009fa}. 

A large boost factor may arise from the non-uniformity of the DM distribution
in the halo. The N-body simulations show, however, that the local clumps of dark
matter density are unlikely to contribute to a large enough
$\rho/\rho_{0}$~\cite{Springel:2008by,Diemand:2008in}.
An other possibility of enhancing  the boost factor is that the DM
annihilation cross section may be velocity-dependent which grows at lower
velocities. The DM annihilation cross section today may be much larger than that
at the time of thermal freeze out, and thus is not constrained by the DM relic
density. Some enhancement mechanisms have been proposed along this line, such
as the Sommerfeld enhancement~\cite{Sommerfeld,Hisano:2002fk,Hisano:2003ec,Cirelli:2007xd,ArkaniHamed:2008qn,Pospelov:2008jd,MarchRussell:2008tu,Iengo:2009ni,Cassel:2009wt}. 
In some non-thermal DM scenarios, the number density of the DM particle can be
enhanced by the out of equilibrium decay of some heavier unstable particles if
the DM particle is among the decay products of the decaying
particle~\cite{Fairbairn:2008fb,Feldman:2009wv}.  The decay of the unstable
particle must take place at very late time. Otherwise the DM particles with
the enhanced number density will annihilate into the Standard Model (SM)
particles again, which washes out the enhancement  effect. 

In this talk, we discuss an alternative origin of the boost factor arising 
from the late time dark matter conversion processes, which requires neither 
the velocity-dependent annihilation cross section nor the decay of unstable
particles~\cite{Liu:2011aa}.  We show that in the scenarios of interacting
multi-component DM, the interactions among the DM components may convert the
heavier DM components into the lighter ones, which is not sensitive to the
details of the conversion interactions.  If the interactions are strong enough
and the DM components are nearly degenerate in mass, the conversion can
enhance the number density of the lighter DM components at late time after the
thermal decoupling. Eventually, the whole DM today in the Universe may
consist of only the lightest DM component with enhanced number density, which
leads to a large boost factor. The scenarios of multi-component DM have
been discuss previously in
Refs.~\cite{Boehm:2003ha,Hur:2007ur,Adibzadeh:2008pe,Feng:2008ya,Zurek:2008qg,Batell:2009vb,Profumo:2009tb,Zhang:2009dd,Gao:2010pg,Feldman:2010wy}.
Note however that the models with simply mixed non-interacting multi-component
DM cannot generate large boost factors.



Let us consider a generic model in which the whole cold DM contains $N$
components $\chi_{i}\ (i=1,\dots,N)$, with masses $m_{i}$ and internal degrees
of freedom $g_{i}$ respectively. The DM components are labeled such that
$m_{i}<m_{j}$ for $i<j$, thus $\chi_{1}$ is the lightest DM particle. We are
interested in the case that $\chi_{i}$ are nearly degenerate in mass, namely
the relative mass differences between $\chi_i$ and $\chi_1$ satisfy
$\varepsilon_{i}\equiv(m_{i}-m_{1})/m_{1}\ll1$. In this case, we shall show
that the interactions between the DM components lead to the DM conversion. 
The thermal evolution of the DM number density normalized to the entropy density
$Y_{i}\equiv n_{i}/s$ with respect to the rescaled temperature $x\equiv
m_{1}/T$ is govern by the following Boltzmann equation
\begin{eqnarray}
&&\frac{dY_{i}(x)}{dx} =
-\frac{\lambda}{x^{2}}
\left[
\langle\sigma_{i}v\rangle(Y_{i}^{2}-Y_{ieq}^{2})
-\sum_{j}\langle\sigma_{ij}v\rangle(Y_{i}^{2}-r_{ij}^{2}Y_{j}^{2})
\right] , 
\label{Boltzmann-eq}
\end{eqnarray}
where $\lambda\equiv x s/H(T)$ is a
combination of $x$, the entropy density $s$ and the Hubble parameter $H(T)$ as a function
of temperature $T$. 
$Y_{ieq}\simeq (g_{i}/s)[m_{i}T/(2\pi)]^{3/2}\exp(-\varepsilon_i x)$ is
the equilibrium number density normalized to entropy density for
non-relativistic particles.  $\langle\sigma_{i}v\rangle$ are the thermally
averaged cross sections multiplied by the DM relative velocity for the process
$\chi_{i}\chi_{i}\to XX'$ with $XX'$ standing for the light SM particles which are
in thermal equilibrium, and $\langle\sigma_{ij}v\rangle$ are the ones for the DM
conversion process $\chi_{i}\chi_{i}\to\chi_{j}\chi_{j}$.
The quantity 
\begin{equation}\label{ratio-r}
r_{ij}(x)\equiv \frac{Y_{ieq}(x)}{Y_{jeq}(x)}
=\left(\frac{g_i}{g_j} \right)
\left(\frac{m_i}{m_j} \right)^{3/2}
\exp[-(\epsilon_i-\epsilon_j)x]
\end{equation}
is the ratio between the two equilibrium number density functions for components  $i$ and $j$. In Eq. (\ref{Boltzmann-eq}) we have assumed kinetic equilibrium.  The first
term in the r.h.s. of Eq.(\ref{Boltzmann-eq}) describes the change of number
density of $\chi_i$ due to the annihilation into the SM particles, and  the
second term describes the change due to the conversion to other DM components.

In the case that the cross section of the conversion process $\langle
\sigma_{ij}v \rangle$ is large enough, the DM particle $\chi_i$ can be kept in
thermal equilibrium with $\chi_j$ for a long time after both $\chi_i$ and
$\chi_j$ have decoupled from the thermal equilibrium with the SM particles.
In this case, the number densities of $\chi_{i,j}$ satisfy a simple relation
\begin{equation}\label{ratio}
\frac{Y_{i}(x)}{Y_{j}(x)}\approx \frac{Y_{ieq}(x)}{Y_{jeq}(x)}=r_{ij}(x) .
\end{equation}   
%
Even when $\chi_{i}$ is in equilibrium with
$\chi_j$ the ratio of the number density $Y_{i}(x)/Y_{j}(x)$ can be quite
different from unity and can vary with temperature. For instance, if $g_i\gg
g_j$ and $0< (\epsilon_i-\epsilon_j) \ll 1$, from Eq. (\ref{ratio-r}) and
(\ref{ratio}) one obtains $Y_{i}(x)\gg Y_{j}(x)$ at the early time when
$(\epsilon_i-\epsilon_j)x \ll 1$. However, at the late time when
$(\epsilon_i-\epsilon_j)x \gg 1$, one gets $Y_{i}(x)\ll Y_{j}(x)$, which is
simply due to the Boltzmann suppression factor $\exp[-(\epsilon_i-\epsilon_j)x]$ in the
expression of $r_{ij}$. Thus the heavier particles can be gradually converted into
lighter ones through this temperature-dependent equilibrium between $\chi_i$
and $\chi_j$.


An interesting limit to consider is that the rates of DM conversion  are large
compared with that of the individual DM annihilation into the SM particles,
i.e.  $\langle\sigma_{ij}v\rangle \gtrsim \langle\sigma_{i}v\rangle$.  In this
limit, after both the DM components have decoupled from the thermal equilibrium
with the SM particles, which take place at a typical temperature $x=x_{dec}\approx25$, the strong
interactions of conversion will maintain an equilibrium between $\chi_i$ and $\chi_j$ 
for a long time
until the rate of the conversion cannot compete with the expansion rate of the
Universe.  Making use of Eq. (\ref{ratio}),  the evolution of the total density
$Y(x)\equiv\sum_{i=1}^{N}Y_{i}(x)$ can be written as
\begin{eqnarray}\label{evolution-sum}
\frac{dY}{dx} & = & 
-\frac{\lambda}{x^{2}}\langle\sigma_{eff}v\rangle\left(Y^{2}-Y_{eq}^{2}\right) ,
\end{eqnarray}
where $\langle\sigma_{eff}v\rangle$ is the effective thermally averaged 
product of DM annihilation cross section and the relative velocity which 
can be written as
\begin{eqnarray}
\langle\sigma_{eff}v\rangle
=
\frac{\sum_{i=1}^{N}w_{i}g_{i}^{2}(1+\varepsilon_{i})^{3}\exp(-2\varepsilon_{i}x)}{g_{eff}^{2}} \langle\sigma_{1}v\rangle ,
\end{eqnarray}
where  $w_{i}\equiv\langle\sigma_{i}v\rangle/\langle\sigma_{1}v\rangle$
is the  annihilation cross section relative to that of the lightest one.
The total equilibrium number density can be written as
\begin{equation}
Y_{eq}\equiv\sum_{i=1}^{N}Y_{ieq}(x)\approx
g_{eff}\left(
  \frac{m_{1}T}{2\pi}
\right)^{3/2} \exp(-x)  ,
\end{equation}
with  an effective degrees of
freedom $g_{eff}=\sum_{i}g_{i}(1+\varepsilon_{i})^{3/2}\exp(-\varepsilon_{i}x)$.  
Note that the conversion terms do not show up explicitly in
Eq. (\ref{evolution-sum}).  Through the conversion processes
$\chi_i\chi_{i}\to \chi_j\chi_{j}$ the slightly heavier components will be
converted into the lighter ones, because the factor $r_{ij}(x)$ is
proportional to $\exp[-(m_i-m_j)/T]$ which suppresses the density of the
heavier components at lower temperature.  If the conversion cross section is
large enough, most of the DM components will be converted into the lightest
$\chi_{1}$ before the interaction of conversion decouples, which may result in
a large enhancement of the relic density of $\chi_1$ and leads to a large
boost factor.

As an example, let us  consider a generic DM model with only two
components. For relatively large conversion cross section $u \equiv \langle
\sigma_{21}v \rangle /\langle \sigma_1 v\rangle \gtrsim 1$, The effective
total cross section is given by
$\langle\sigma_{eff}v\rangle=\langle\sigma_{1}v\rangle[1+w g^{2}\exp(-2\varepsilon x)]/[1+g\exp(-\varepsilon x)]^{2}]$, 
where $w\equiv w_2 $, $g\equiv g_{2}/g_{1}$ and
$\varepsilon\equiv\varepsilon_{2}$.  Because of the $x$-dependence in
$\langle\sigma_{eff}v\rangle$, the thermal evolution of $Y(x)$ differs
significantly from that of the standard WIMP. In the case that $\chi_2$ has
large degrees of freedom but a small annihilation cross section,
namely $g\gg1$, $w\ll1$ and $wg^{2}\ll1$, the thermal evolution of the total
density $Y$ can be simplified. 
%
%
The thermal evolution of the total number density can be roughly divided into four stages:  
i) At high temperature region where $3 \lesssim x\ll x_{dec}$, both the DM
components are in thermal equilibrium with the SM particles.  $Y_{i}(x)$ must
closely track $Y_{ieq}(x)$ which decrease exponentially as $x$ increases.
However,  since $g\gg 1$ and
$\epsilon \ll 1$, the number density of $\chi_2$ is much higher than
that of $\chi_1$, i.e. $Y_2(x) \gg Y_1(x)$.
ii) When the temperature goes down and $x$ is close to the decoupling point
$x_{dec}$, both the DM components start to decouple from the thermal
equilibrium. In the region $x_{dec}\lesssim x\ll 1/\varepsilon$,
$\langle\sigma_{eff}v\rangle$ is nearly a constant and
$\langle\sigma_{eff}v\rangle\approx\langle\sigma_{1}v\rangle/(1+g)^{2}\ll\langle\sigma_{1}v\rangle$,
the total density $Y(x)$ behaves just like that of an ordinary WIMP which
converges quickly to $Y(x)\approx x_{dec}/(\lambda \langle \sigma_1 v
\rangle)$.
iii) As $x$ continues growing, the suppression factor $\exp(-\varepsilon x)$
in $\langle\sigma_{eff}v\rangle$ becomes relevant. The value of
$\langle\sigma_{eff}v\rangle$ grows rapidly especially after $x$ reaches the
point $\varepsilon x\approx\mathcal{O}(1)$, which leads to the further
reduction of $Y(x)$. In this stage, although both $\chi_{1,2}$ have decoupled
from the thermal equilibrium with the SM particles. The strong conversion
interaction $\chi_2\chi_2 \leftrightarrow \chi_1\chi_1$ maintains an
equilibrium between the two DM components. According to Eq. (\ref{ratio}), the
relative number density $Y_2(x)/Y_1(x)$ decreases with $x$ increasing, which
corresponds to the conversion from the heavier DM component into the lighter
one.  At the point $x_c= (1/\varepsilon)\ln g$ one has
$Y_2(x)\approx Y_1(x)$. 
For the region $x>x_{dec}$ and $x$ is not close to
$x_c$, because of $Y_{eq}(x)\ll Y(x)$ and $g\exp(-\varepsilon x)\gg 1$,
the total number density  can be analytically integrated out, and 
$Y(x)$ in this region can be approximated by
\begin{equation}
Y(x)\approx \frac{g^2 x_{dec}}{\lambda \langle \sigma_1 v\rangle }
\left[ 1+
\left(\frac{ x_{dec}}{x}\right)\frac{\exp(2\varepsilon x )}{2\varepsilon x}
\right]^{-1}  .
\label{approx-solution}
\end{equation}
iv) When $x$ becomes very large $\varepsilon x\gg\mathcal{O}(1)$ ,
$\langle\sigma_{eff}v\rangle$ quickly approaches $\langle\sigma_{1}v\rangle$,
and becomes independent of $x$ again. The evolution of $Y(x)$ in this region can be
obtained by a simple integration as it was done in the stage ii). The solution of $Y(x)$
shows a second decoupling.  Finally when the conversion rate cannot compete
with the expansion rate of the Universe at some point $x_F$ corresponding to $
sY_2\langle \sigma_{21} v\rangle/H \approx 1$, both $Y_1(x)$ and $Y_2(x)$ remain
unchanged as relics. The whole DM can be dominated by $\chi_{1}$ if
the conversion is efficient enough.

By matching the analytic solutions of $Y(x)$ in different regions near the points
$x_{dec}$ and $x_c$, and requiring that the final total relic density is
equivalent to the observed $\Omega_{CDM} h^2\approx 0.11$, we obtain the following
approximate expression of the boost factor
\begin{equation}
B\approx g^2 \left[1+ \left(\frac{x_{dec}}{x_c}\right) \left(\frac{\exp(2\varepsilon x_c)}{2\varepsilon x_c}+g^2 \right) \right]^{-1} .
\label{boostFacEq}
\end{equation}
As expected, the enhancement essentially comes from the conversion of the
degrees of freedom. Thus the maximum enhancement is $g^2$. The two 
terms in the r.h.s of the above equation correspond to the reduction of $Y(x)$
during the late time conversion stages.  For large enough $g$, the boost
factor can be approximated by $B\approx g^2/(1+\varepsilon g^2 x_{dec}/\ln g
)$. In order to have a large boost factor, a small $\varepsilon \ll \ln g/(g^2
x_{dec})$ is also required. As shown in Eq. (\ref{boostFacEq}) the boost factor is
not sensitive to the exact values of the cross sections as long as the
conditions $w \ll 1$ and $u\gg 1$ are satisfied.

\begin{figure}[htb]
\begin{center}
\includegraphics[width=0.5\columnwidth]{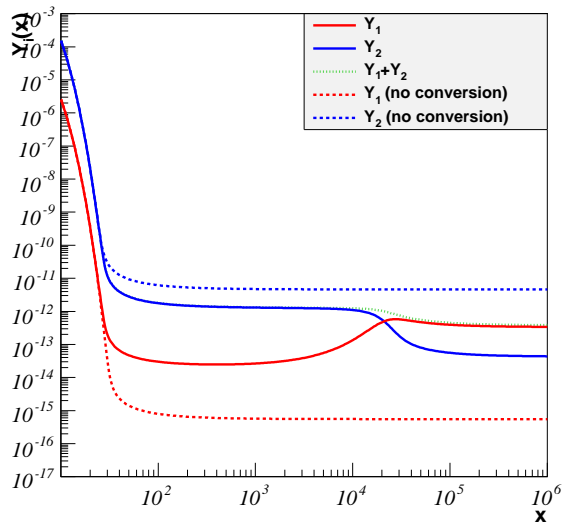}
\end{center}
\caption{
Thermal evolution of the number densities $Y_1(x)$ (red solid) and $Y_2(x)$ (blue solid) 
with respect to $x$. The  solid (dashed) curves correspond  to the case with (without) DM 
conversions. The green dotted curve  corresponds to the sum of $Y_1$ and $Y_2$,  
for parameters $g=60$, $m_1=1$TeV,
 $\varepsilon =2\times 10^{-4}$, $w=10^{-4}$ and $u=10$ respectively.
 \label{fig:Time-evolution}
}

\end{figure}

We numerically calculate the thermal evolution of $Y_i(x)$ and the boost
factor without using approximations for a generic two-component DM model. The
results for $w=10^{-4}$, $u=10$ and $\varepsilon=2\times 10^{-4}$ is shown in
Fig.~\ref{fig:Time-evolution}.  The value of $\langle\sigma_{2}v\rangle$ is
adjusted such that the final total DM relic abundance is always equal to the
observed value $\Omega_{CDM}h^{2}$.  The mass of the light DM particle is set
to $m_{1}=1$ TeV. For an illustration the ratio between the internal degrees
of freedom is set to be large $g=60$.
From the figure, the four stages of the thermal evolution of $Y(x)$ as well as
the crossing point can be clearly seen.  The crossing point at $x=x_c \approx
2\times 10^{-4}$ indicates the time when the number density of $\chi_{1}$
start to surpass that of $\chi_{2}$ and eventually dominant the whole DM relic
density. In this parameter set a large boost factor
$B\approx\langle\sigma_{1}v\rangle/\langle\sigma v\rangle_F\approx 585 $ is
obtained which is in a remarkable agreement with Eq. (\ref{boostFacEq}) with
error less than $\sim 5\%$.  For a comparison, in
Fig. \ref{fig:Time-evolution} we also show the cases without conversions.
%


The whole DM in the universe necessarily contains multiple components, as the lightest
active neutrino  already contributes to a small fraction of the 
DM relic density.  It is easy to construct models with more stable neutrinos or neutrinos
with lifetime longer than that of the universe. For instance,
in fourth generation models with right-handed neutrinos, extra stable neutrinos
may be the keV scale sterlile neutrinos and the heavy Majorana neutrinos which
are stable due to additional symmetries\cite{Zhou:2011fr}.
For models with multiple DM components, it is possible that there exists interactions among
the DM components which may lead to the conversions among  them.  In this talk we
consider a simple interacting two-component DM model by adding to the standard
model (SM) with two SM gauge singlet fermionic DM particles $\chi_{1,2}$. The
particles $\chi_{1,2}$ are charged under a local $U(1)$ symmetry which is
broken spontaneously by the vacuum expectation value (VEV) of a scalar field
$\phi$.  The corresponding massive gauge boson is
denoted by $A$ which may cause the reaction $\bar{\chi_2}\chi_2
\leftrightarrow \bar{\chi_1}\chi_1$.  The stability  of $\chi_{1,2}$ is
protected by two different global $U(1)$ number symmetries.
%
An SM gauge singlet pseudo-scalar $\eta$ is introduced as a messenger field
which couples to both the dark sector and the SM sector.
%
In order to have the leptophilic nature of DM annihilation, we also introduce
an SM $SU(2)_{L}$ triplet field $\Delta$ with the SM quantum number $(1,3,1)$
and flavor contents $\Delta=(\delta^{++},\delta^{+},\delta^{0})$.  The triplet
carries the quantum number $B-L$=2 such that it can couple to the SM
left-handed leptons $\ell_L$ through Yukawa interactions $\bar{\ell}^c_L
\Delta \ell_L$, but cannot couple to quarks directly.  The VEV of the triplet
has to be very small around eV scale, which is required by the smallness of
the neutrino masses. As a consequence, the couplings between one triplet and
two SM gauge bosons such as $\delta^{\pm\pm}W^{\mp}W^{\mp}$,
$\delta^{\pm}W^{\mp}Z^0$ and $\delta^{0}Z^0Z^0$ are strongly suppressed as
they are all proportional to the VEV of the triplet, which makes it difficult
for the triplet to decay even indirectly into quarks through SM gauge bosons
\cite{Gogoladze:2009gi,Guo:2011zze,Guo:2010sy,Guo:2010vy,Guo:2008si,Wu:2007kt}. 
If $\eta$ has a stronger coupling to
$\Delta$ than that to the SM Higgs boson $H$ and $\phi$ then the annihilation
products of the dark matter particles $\chi_{1,2}$ will be mostly leptons.

%
The Lagrangian of the model can be written as $\mathcal{L}=\mathcal{L}_{SM}+\mathcal{L}_1$
The new interactions in $\mathcal{L}_1$ which are relevant to the DM annihilation and conversion are 
given by
\begin{eqnarray}\label{lagrangian}
\mathcal{L}_1 
&\supset &
\bar{\chi}_i (i\slashed D-m_i )\chi_i
+(D_\mu \phi)^\dagger (D^\mu \phi)- m_\phi^2 \phi^\dagger\phi  
\nonumber\\
&&+\frac{1}{2}\partial_\mu\eta\partial^\mu\eta-\frac{1}{2} m_\eta^2 \eta^2 -y_i \bar{\chi}_i i\gamma_5 \eta \chi_i -y_\ell \bar{\ell}^c_L \Delta \ell_L+\mbox{h.c}
\nonumber\\
&& -(\mu \eta+\xi\eta^2 )
\left[
  \mbox{Tr}(\Delta^\dagger \Delta )+\kappa (H^\dagger H)+\zeta (\phi^\dagger\phi)
\right],  \;\; (i=1,2)
\end{eqnarray}
Note that $\phi$ and $\eta$ do not directly couple to the
SM fermions. 
After the spontaneous symmetry breaking in $V(\phi)$, 
the scalar $\phi$ obtains a nonzero VEV $\langle \phi \rangle=v_\phi/\sqrt{2}$  
which generates the mass  of  the gauge boson  $m_A=g_A v_\phi$. 
At the tree level, the three components of the triplet $\delta^{++},\delta^{+}$
and $\delta^0$ are degenerate in mass, i.e.  
$m_{\delta^{++}}=m_{\delta^{+}}=m_{\delta^{+}}\equiv m_\Delta$. 

We assume that $\chi_2$ has large internal degrees of freedom relative to that
of $\chi_1$, i.e., $g_2\gg g_1$, which can be realized if $\chi_2$ belongs to
a multiplet of the product of some global nonabelian groups.  For instance
$g_2=4\tilde{g}_2$ with $\tilde{g}_2=$16, 8, and 4 if it belongs to the spinor
representation of a single group of $SO(8)$, $SO(6)$ and $SO(4)$ respectively.
When $\chi_2$ belongs to a representation of the product of these groups, its
internal degrees of freedom can be very large. 

At the early time when the temperature of the Universe is high enough, the
triplet $\Delta$ can be kept in thermal equilibrium with SM particles through
the SM gauge interactions.  The DM particles $\chi_i$ are in thermal
equilibrium by annihilating into the triplet through the intermediate particle
$\eta$.  
The annihilation
$\bar{\chi}_2\chi_2\to\eta^*\to \delta^{\pm\pm}\delta^{\mp\mp},
\delta^{\pm}\delta^{\mp}, \delta^{0}\delta^{0*} $ is an $s$-wave process which
is the dominant contribution . 
The ratio of the two annihilation cross sections is $w=(y_2/y_1)^2(g_1/g_2)$.
It is easy to get a very small $w$ provided that $y_2 \ll y_1$ and $g_1 \ll
g_2$. In order to have a large enough $\langle \sigma_1 v \rangle \gg \langle
\sigma v\rangle_F$ the product of the coupling constants $y_1 \mu$ must be
large enough, or the squared mass of $\eta$ is close to $s$.
The cross section of the conversion process $\bar{\chi}_2\chi_2\to A^* \to
\bar{\chi}_1\chi_1$ is 
suppress by $g_1/g_2$ and also the phase space factor
$\sqrt{1-4m_1^2/s}$ when $s$ is close to $4m_2^2$ at the vary late time of the
thermal evolution. However, the cross section can be greatly enhanced if $m_A$ is
close to a resonance when the relation $s \simeq m_A^2$ is satisfied.  In the
numerical calculations, we find that for the following selected  parameters:
$m_1=1$TeV, $\epsilon=1\times 10^{-4}$, $g_1=1$, $g_2=60$, $m_{\Delta}=500$
GeV, $m_\eta=1.5$ TeV, $m_A=2.02$ TeV, $y_1=3$, $y_2=0.07$, $\mu/m_1=3$, and
$g_A=2.5$, the following ratio of the cross section can be obtained
$$
w\simeq 1\times 10^{-5}, \ u\simeq 0.5, \
\mbox{and} \  \langle \sigma_1 v \rangle /\langle \sigma v\rangle_F \simeq 500 .
$$
In this parameter set the relative mass difference between $m_A$ and $2m_2$ is around
$1\%$. 
The  corresponding boost factor is $B\sim 500$, which is large enough to account for the
PAMELA data for the dark matter mass around TeV.  

In summary, We have considered an alternative mechanism for obtaining
boost factors from DM conversions  which does not require the
velocity-dependent annihilation cross section or the decay of unstable
particles.  We have shown that if the whole DM is composed of multiple
components, the relic density of each DM component may not necessarily be
inversely proportional to its own annihilation cross section. We demonstrate
the possibility that the number density of the lightest DM component can get
enhanced in late time through DM conversation processes, and finally dominates
the whole relic abundance, which corresponds to a  boost factor needed to
explain the excesses in cosmic-ray positron and electrons reported by the
recent experiments.
\section*{\ack}
This work is supported in part by the National Basic Research Program
of China (973 Program) under Grants No. 2010CB833000; the National
Nature Science Foundation of China (NSFC) under Grants No. 10975170,
No. 10821504 and No. 10905084; and the Project of Knowledge Innovation
Program (PKIP) of the Chinese Academy of Science.
\section*{References}

\providecommand{\newblock}{}

\end{document}